%% file: template.tex
\documentclass{Interspeech2024}

\interspeechcameraready

\title{The Whole Is Bigger Than the Sum of Its Parts: Modeling Individual Annotators to Capture Emotional Variability}

\name[affiliation={1}]{James}{Tavernor}
\name[affiliation={1}]{Yara}{El-Tawil}
\name[affiliation={1}]{Emily}{Mower Provost}

\address{
 $^1$University of Michigan, USA}
\email{tavernor@umich.edu, yeltawil@umich.edu, emilykmp@umich.edu}


\usepackage{lscape}
\usepackage{subcaption}
\usepackage{algorithm}
\usepackage{algpseudocode}
\usepackage{multirow}
\def\version{new} 
\def\both{both}
\def\old{old}

\usepackage[normalem]{ulem}
\newcommand{\swap}[2]{\ifx\version\both{\color{red}\sout{#1} \color{blue}#2}\else\ifx\version\old#1\else#2\fi\fi}
\begin{document}

\maketitle

\begin{abstract}

Emotion expression and perception are nuanced, complex, and highly subjective processes. When multiple annotators label emotional data, the resulting labels contain high variability. Most speech emotion recognition tasks address this by averaging annotator labels as ground truth. However, this process omits the nuance of emotion and inter-annotator variability, which are important signals to capture. Previous work has attempted to learn distributions to capture emotion variability, but these methods also lose information about the individual annotators. We address these limitations by learning to predict individual annotators and by introducing a novel method to create distributions from continuous model outputs that permit the learning of emotion distributions during model training.  We show that this combined approach can result in emotion distributions that are more accurate than those seen in prior work, in both within- and cross-corpus settings.  
\end{abstract}

\section{Introduction}
Expressions of emotion are nuanced and complex, and people perceive these expressions differently, adding to the complexity. Most emotion recognition models overlook this nuance~\cite{labat2022variation}. This is because most Speech Emotion Recognition (SER) datasets and tasks present the ground truth as a single label, which is the average of multiple annotations. In this work, we present novel approaches to both accurately learn the perceptions of individual annotators and aggregate these estimates to create distributions of annotator perception. In this way, the model retains information about individual annotator predictions while still being able to summarize the information accurately as a two-dimensional (2D) distribution. 





Prior work has investigated methods to retain information about variability and uncertainty. Research has included the prediction of measures such as unbiased annotator standard deviation~\cite{han2017hard,prabhu2021end}, the embedding of individual annotators to improve performance on the aggregated ground truth, with some investigation into how well the model annotator uncertainty correlates with real uncertainty~\cite{davani2022dealing,kocon2021,chou2020learning}, and the prediction of the distribution of annotations over a given utterance~\cite{zhang2017predicting}. Yet, gaps remain. Methods that summarize model information or predict uncertainty lose fine-grained information about individual annotators. On the other hand, methods that seek to learn annotators primarily do so to improve performance on the aggregated ground truth or investigate much smaller numbers of annotators than are generally used in these datasets.

We present a novel approach that predicts the annotations of individuals and includes a new differentiable method to automatically learn distributions similar to~\cite{zhang2017predicting}, enabling the modeling of individual variation and the retention of the ability to summarize annotators. The model training involves an interleaved approach, alternating between different tasks: learning individual annotators and learning a distribution. We learn individual annotators by training a multi-task (MT) model to predict each annotator in the training set across the dimensions of valence and activation. We learn a distribution by upsampling the observations from the MT model and using Kernel Density Estimation (KDE) to produce a summarization of the model output as a distribution. We introduce differentiable KDE into the model training process to enable the use of gradient descent. 

We present both within- and cross-corpus investigations. Within-corpus, we find that a model trained with the interleaved tasks of individual annotator perception and distribution learning can outperform a method that learns to predict the distribution alone~\cite{zhang2017predicting}, in terms of both the performance on consensus labels and the accuracy of the distribution itself, while providing individual annotations as well. We further show that the output of the annotator-specific models (trained only on annotator prediction) can be post-processed to create a distribution, rather than learning a distribution during model training, that outperforms the prior work of~\cite{zhang2017predicting}. In this case, an extra step is involved in which the output of the annotator-specific models is transformed into a distribution using either KDE as in~\cite{zhang2017predicting} or using the differentiable KDE method presented in this work. We find that using differentiable KDE leads to significantly improved performance, even when only used in post-processing, pointing to the efficacy of this approach for either model learning or post-hoc output summarization. Cross-corpus, we demonstrate that annotator-specific models can be used zero-shot without knowledge about the annotators that labeled the new datasets. We find that the presented approach outperforms a distribution-only method across metrics that capture individual annotators and the accuracy of a given distribution in most cases. Future work will focus on investigating individual characteristics of annotators (e.g., personality) and how this information can also be considered when learning annotator-specific perception.

\section{Related Work}
\swap{}{Previous work has developed soft-label methods that use multiple annotators per label. Dang et al.~\cite{req2} use multi-rater Gaussian Mixture Regression to make temporal emotion predictions for a fixed set of consistent evaluators in their target dataset. Other approaches have captured both the uncertainty in annotator labels and model uncertainty~\cite{req1}. However, a gap remains at the intersection of predicting individual annotations for a variable number of annotators.

}
Instead, we build on the label processing method developed in previous work by Zhang et al.~\cite{zhang2017predicting}, which incorporated inter-annotator variance into machine learning models by creating new ground truth labels that incorporate this knowledge~\cite{zhang2017predicting}. 
They upsampled existing annotations by selecting random subsets of annotators for each utterance and took the mean across those annotations. They added random noise to the resulting means, such that $x\ noise\sim U(-\frac{std(x)}{2}, \frac{std(x)}{2})$\footnote{We also add $\epsilon=1E^{-12}$ to this value to account for cases where standard deviation is 0.}, where $x$ is activation or valence, $std$ indicates the standard deviation of the annotator ratings for that utterance, and $U$ is the uniform distribution. Kernel Density Estimation (KDE) via Diffusion was then calculated over the upsampled observations. They divided the KDE output grid into $N$ bins for each dimension and took the mean over the KDE samples inside each bin. They converted this grid to a probability distribution by normalizing over the means. The authors investigated $N=2$ and $N=4$. The KDE step was essential to remove sensitivity to where boundaries were drawn. The authors then trained a model to predict these binned distributions. However, in this approach, the model loses information about individual annotators. Additionally, because the approach is not differentiable, it cannot be included in model training. We present an approach with a differentiable component that permits learning a binned distribution, implemented using sigmoid-based soft operations.

Previous work has investigated the prediction of individual annotators on subjective tasks such as emotion recognition and hate speech~\cite{davani2022dealing,kocon2021,9666407}. Davani et al.~introduced an encoder-based model with separate classification heads for each annotator. They trained this model for a binary categorical text emotion recognition task using a dataset that contained 82 annotators. At test time, they aggregated the individual annotator predictions and found that their model outperformed a baseline trained on majority ground truth labels. However, the performance of individual annotators was not discussed. Further, a limitation of this work is that many SER datasets include over 82 annotators, and the authors acknowledge that it would be too computationally expensive to train a model with separate heads for large numbers of annotators. Previous work has shown that clustering similar annotators can mitigate problems with large numbers of annotators~\cite{10417135}. However, clustering annotators loses information about individual ratings. In our work, we enable only the relevant heads per batch, making training with a large number of annotators more computationally feasible.

An alternative approach to learning individual annotators is through annotator embeddings~\cite{kocon2021}. Prior work from Kocoń et al. demonstrates that annotator-specific embeddings can be used to personalize model predictions and capture the bias of individual annotators. They introduced four methods for encoding annotator information into the model, including a one-hot annotator embedding. This embedding was a one-hot encoded vector of annotator ID that was concatenated to the model input. They found that this led to improved text-based emotion predictions but were focused on a consensus model rather than an individual-specific model. We use the one-hot model and investigate if the model can learn individual annotators.

\section{Experiments}
\subsection{Data setup}
We use the MSP-Improv dataset for training and testing. It was labeled using crowdsourcing and has a relatively large number of evaluations per utterance~\cite{zhang2017predicting,improv}. Additionally, we use the IEMOCAP, MSP-Podcast, and MuSE datasets to evaluate the cross-corpus results of each method.

\textbf{MSP-Improv} is an SER dataset consisting of acted improvised dialogue designed to evoke certain emotions~\cite{improv}. The dataset has 12 speakers evenly split between male and female actors across six sessions. We select a speaker-independent data split such that all annotators in the validation and test set have evaluated at least one utterance in the training set. Annotators will be present in the training set that do not appear in the validation or test set (for example, when the annotator annotated less than three samples). The resulting train, validation, and test split size is 5,851, 1,287, and 1,300 utterances, respectively\footnote{The code to create the data splits can be found at https://github.com/chailab-umich/ModelingIndividualEvaluators.}. The training set was evaluated by 1,434 individual crowdsourced annotators, with each sample receiving between 5 and 50 annotations (mean of 7.2). A subset of these annotators evaluated the validation and test set (1,305 and 1,197, respectively). Both validation and test set samples have between 5 and 37 evaluations per sample, with a mean of 7.3 and 7.6 annotators.
In few samples (28) the same annotator has annotated more than once. In these cases we have averaged their annotations into one evaluation, and adjusted the mean ground truth for these samples. 

The \textbf{IEMOCAP} dataset contains five dialogue sessions containing scripted and improvised interactions between two actors. There is one female and one male actor in each conversation~\cite{iemocap}. We remove utterances where individual annotations were partially missing or any annotator evaluations were not within the labeling range described in the data collection. After processing, the dataset consists of 9,999 samples. Six annotators labeled the dataset with an average of 2.13 annotators per sample. We test on the full dataset.

\textbf{MSP-Podcast} is a dataset of speech taken from podcasts and then labeled~\cite{podcast}. We use the predefined splits and evaluate on test\_set\_1, which is comprised of 13,911 utterances and contains 9570 individual annotators. Each utterance was evaluated by 6.9 crowdsourced annotators on average. We use \texttt{release 1.8}, which does not contain transcripts, so we use Microsoft Azure automatic speech recognition to generate them. 

\textbf{MuSE} is a dataset of 28 college students recorded in two 45-minute sessions each, responding to emotional stimuli. One session was when the students were affected by an external stressor, and the other was without the stressor~\cite{jaiswal-etal-2020-muse}. Students were recorded using a lapel microphone. Crowdsourced annotators evaluated each utterance. There are 2,584 utterances comprised of 1,385 stressed and 1,199 non-stressed samples. The dataset provides labels annotated with or without context; we use the labels from the 160 individual annotators who labeled without context. Each sample was evaluated between 7 and 9 annotators, with 8 on average. 

\textbf{Dataset Preprocessing}
We process all datasets in the same way. We use min-max scaling on the annotator and consensus labels for activation and valence to restrict labels to the $[-1, 1]$ range. We then use KDE to generate a 2D ground truth probability distributions as in~\cite{zhang2017predicting}. We use a KDE grid size of 512 as we assume this will be sufficiently large to ensure the probability is insensitive to the grid boundaries.

\subsection{Model Architecture}
\begin{figure*}[t]
 \begin{subfigure}{0.30\textwidth}
 \includegraphics[width=\linewidth]{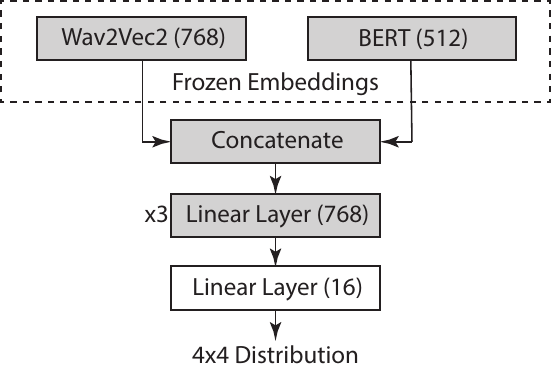}
 \caption{Baseline model} \label{fig:1a}
 \end{subfigure}%
 \hspace*{\fill} 
 \begin{subfigure}{0.30\textwidth}
 \includegraphics[width=\linewidth]{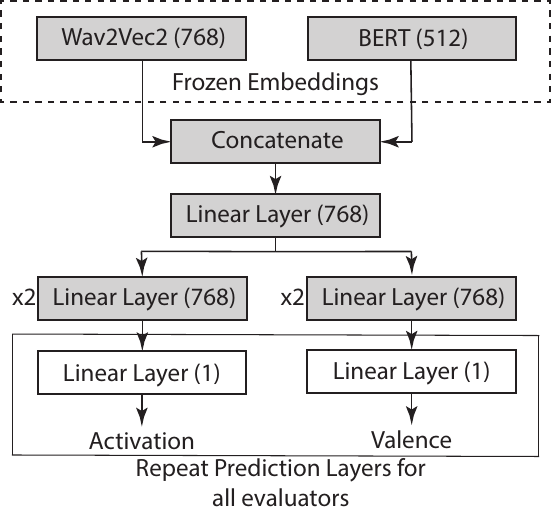}
 \caption{MT Model} \label{fig:1b}
 \end{subfigure}%
 \hspace*{\fill} 
 \begin{subfigure}{0.30\textwidth}
 \includegraphics[width=\linewidth]{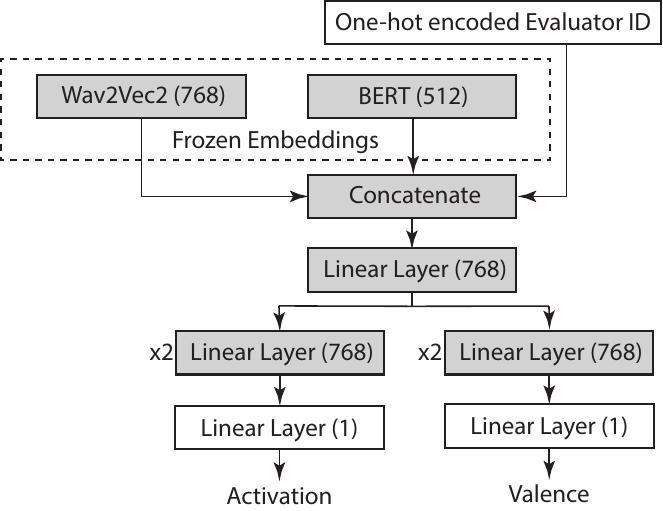}
 \caption{One-hot Model} \label{fig:1c}
 \end{subfigure}
\caption{Model Architectures. Layers in gray are the common architecture between models. In (b) and (c) the last two common layers are duplicated as they split the model to two predictions.}\label{fig:model-arch}
\vspace{-18pt}
\end{figure*}

We present three models: a baseline, a MT model, and a one-hot model, all of which share the same base architecture but have different output head architectures (Figure~\ref{fig:model-arch}). The model input includes the frozen mean-pooled final layers of Wav2Vec2~\cite{baevski2020wav2vec} and BERT~\cite{Devlin2019BERTPO} CLS embeddings as these have shown effectiveness in SER applications~\cite{tavernor23_interspeech,wagner2023dawn}. We apply dropout with probability $0.2$ and concatenate the embeddings. The concatenated embedding is passed through a single linear layer of size $256$ with ReLU activation. For each prediction (distribution, activation, or valence), the input will pass through two linear layers of size $256$ with ReLU activations. 

The \textbf{baseline} model directly learns the generated KDE distribution (as in~\cite{zhang2017predicting}), having a final linear layer output of 16 logits for the 4x4 discretized KDE distribution prediction case. The \textbf{MT} model has separate prediction layers for each annotator as in~\cite{Devlin2019BERTPO}. Each annotator's continuous prediction of activation and valence is made via a linear layer with an output size of 1. We use the \textbf{one-hot} method, previously used for text emotion recognition~\cite{kocon2021learning}. We use the same architecture as in the MT case but with only one annotator prediction head. The annotator ID is one-hot encoded and concatenated to the Wav2Vec2 and BERT embeddings on the model input. When training within corpus, we reduce computation cost by making predictions only for annotators in the batch input to the model.

\subsection{Training Tasks}
In this section, we define three different training tasks. The Baseline model is trained with the Baseline task. MT and one-hot models are trained by interleaving Tasks 1 and 2 (Task 1+2), defined below. We use stochastic gradient descent with a learning rate of $0.001$, with a learning rate scheduler that adjusts the learning rate by a factor of $0.1$ after five epochs of no reduction in validation metrics. We train models until early stopping triggers with a patience of 10 with a minimum of 30 epochs. Each model trains with a batch size of 32. For all methods we use the relevant task's validation losses. 

\textbf{Baseline}: We predict the flattened 2D distribution and use cross-entropy loss of the 16 logits output against the flattened 2D generated ground truth probability distributions~\cite{zhang2017predicting}.

\textbf{Task 1 - Annotator Training}: 
We train annotator-specific predictions using the individual annotator ground truth. We use Lin's Concordance Correlation Coefficient (CCC) loss as our loss function since it better models dimensional attributes than other regression losses~\cite{atmaja2021evaluation}.
The sets $act$, and $val$ contain the ground truth labels from all annotators in a training batch. The sets $m_{act}$ and $m_{val}$ contain the model's estimates of these labels. The loss is $2 - CCC(m_{act}, act) - CCC(m_{val}, val)$.




\textbf{Task 2 - \textit{DiffKDE}}: 
We learn the probability distribution using the KDE-generated ground truth labels. The model must produce a probability distribution from the model's activation and valence predictions. However, the KDE method outlined by Zhang et al.~in~\cite{zhang2017predicting} is not immediately usable. KDE via diffusion starts with a histogram~\cite{10.1214/10-AOS799}. For each annotation we must know if it is in a particular bin to increment the bin's histogram count. This operation is a binary operation and not differentiable. We introduce a differentiable approximation to this problem by instead calculating a confidence value that a given annotation is within a given bin. We modify an existing one-dimensional (1D) soft-histogram\footnote{https://discuss.pytorch.org/t/differentiable-torch-histc/25865/4}, as below, for the 2D data. 

We use 64 bins for \textit{DiffKDE}\footnote{Note: smaller than 512 (used to generate target labels) for speed}. We first calculate the 1D center of each bin in the range $-1$ to $1$. For each of the $n$ annotations of activation, we subtract the center of each bin from the annotation, resulting in a $64$ size vector, which we call $x$. \swap{If the annotation is closest to the center of bin $i$, the closest element in $x$ to 0 will be $x_i$. }{}The contribution to the $64$ bins will then be calculated using an element-wise $sigmoid$ on this vector, $sigmoid(\sigma*(x+\frac{\delta}{2}))-sigmoid(\sigma*(x-\frac{\delta}{2}))$. The gradient of $sigmoid$ is largest at zero, for values of $x$ far from 0, the $\frac{\delta}{2}$ term has less effect, and the bin value is close to 0. This function is maximized for values of $x$ close to 0. We repeat this for valence to get two $n\times 64$ matrices for activation and valence. 

In the equation, $\sigma$ is a scaling parameter; the larger the value, the more sharp the histogram is, and $\delta$ is the bin size. Since our data is in the range $-1$ to $1$, and we use 64 bins, $\delta = \frac{2}{64}$. We then matrix multiply these two $n\times 64$ matrices by transposing one to get a 2D ($64\times 64$) matrix. We then normalize to get a final $4\times 4$ probability distribution as in~\cite{zhang2017predicting}. There is a tradeoff where too large of a $\sigma$ may lead to vanishing gradients, but too low may result in undersaturation~\cite{8843652}. As such, we set $\sigma$ relatively small at $8$; lower values did not reduce loss. Future work could investigate the impact of modifying the $\sigma$ parameter. The generation of probabilities in \textit{DiffKDE} is done in \texttt{float16}\footnote{We use \texttt{float64} during validation and testing for KDE accuracy.} as it significantly speeds up calculations.

We base our work off an existing KDE via Diffusion library\footnote{https://pypi.org/project/KDE-diffusion/}, which we modify to use PyTorch and the soft histogram method from the previous paragraph. All code is available on our GitHub page\footnote{https://github.com/chailab-umich/ModelingIndividualEvaluators}. This enables \textit{DiffKDE} to be run on GPUs and parallelized into batches. \textit{DiffKDE} Loss is the Cross-Entropy loss\footnote{After normalization, we add $\epsilon=1E^{-8}$ to avoid taking $\log$ of 0.} of the \textit{DiffKDE} output, compared with the generated ground-truth 2D labels.



\subsection{Evaluation Metrics}
\vspace{-3pt}
We first evaluate the ability of the proposed approaches to learn continuous predictions and then the ability of the system to learn distributions. The baseline cannot directly produce continuous ratings, while the proposed approaches can. In order to provide a fair comparison, we generate consensus predictions across all methods in the same manner: we sum along the activation/valence dimensions and then multiply this sum with $[-1,-0.5,0.5,1]$. We use CCC to measure the systems' ability to predict individual annotators' labels (note: we cannot evaluate the baseline for this task). Next, we evaluate the consensus predictions by comparing them to the averaged ground truth using CCC. Finally, we measure the differences between the learned and ground truth probability distributions using Total Variation Distance (TVD), and Jensen-Shannon Divergence (JSD)~\cite{zhang2017predicting}\footnote{We use natural logarithm for JSD instead of $\log_2$}. Test results are reported over five seeds. Significance asserted at a 5\% confidence on a paired two-sided t-test.

\section{Results}

\input{tables/improv_results}

\subsection{MSP-Improv Results}
\vspace{-3pt}
The MT approach predicts annotator-specific activation more accurately than the one-hot model ($0.629 \pm 0.002$ vs. $0.349 \pm 0.015$, respectively) while the one-hot model has stronger performance for valence ($0.393 \pm 0.006$ vs. $0.429 \pm 0.007$, respectively). The consensus output for both the MT and one-hot approaches show significant improvements in activation CCC compared to the baseline. 
In contrast, only the one-hot method significantly improves valence. Overall, we find that the MT model learns more accurate distributions compared to the baseline when using the soft-histogram across both metrics, showing signficant improvement over the baseline for TVD. 
The one-hot method has comparable TVD and statistically significantly worse JSD than the baseline. 
See Table~\ref{tab:improv-results} for more details.




\subsection{Ablation Results}
\vspace{-3pt}
We investigate the importance of the interleaved training tasks for learning the distributions. In the previous experiments, we used \textit{DiffKDE} during training and testing (Task 1+2). When using Task 1 alone, no distribution is used during training, so the best method to build the distribution is uncertain. We generated results for Task 1 alone using both \textit{KDE} and \textit{DiffKDE} to generate distributions. We find that when using \textit{DiffKDE} there is a significant performance increase for TVD (0.553±0.003 to 0.500±0.003) and JSD (0.265±0.002 to 0.211±0.002). This is very similar to the performance of Task 1+2 (Table~\ref{tab:improv-results}).


\subsection{Cross-Corpus Results}
\vspace{-3pt}
\input{tables/cc_results}
In a cross-corpus (zero-shot) context, the model does not have information about all annotators in advance. Therefore, we use all annotator predictions from the model. We use the MT approach as it has generally outperformed one-hot models. 
The MT models excel in cross-corpus performance and significantly outperform the baseline in all probability distribution measures (TVD and JSD) on all datasets. Additionally, we find statistically significant increases in Activation CCC performance on the IEMOCAP and MuSE datasets for both the annotator-only trained model (Task 1) and the interleaved tasks trained model (Task 1+2). The outlier is Valence CCC, which generally decreases compared to the baseline. See Table~\ref{tab:cc-results}. 

The MT models generally struggled with the valence dimension, showing significant decreases compared to the baseline. Given that we are using all annotators for zero-shot test time, it is likely many annotator predictions that did not learn valence well have influenced the valence dimension negatively. Ultimately, we believe that using all annotators as we have done in a zero-shot setting is not an upper bound for performance of these models. Instead, selecting a subset of trained annotators may significantly increase performance in the zero-shot setting.

\section{Conclusion}
Learning individual annotators is challenging. The model must learn a very large number of annotators across both the dimensions of activation and valence. We have presented an approach that accurately predicts individual annotators and a differentiable KDE operation that can be applied to a multi-task annotator models to produce distributions more accurately than using KDE to generate the distributions.
We find that a multi-task model sufficiently learns the individual annotators to produce a probability distribution that outperforms methods that only learn distributions while retaining information about individual annotators.
Furthermore, we have found significant improvement in multiple zero-shot settings when using the multi-task model over the baseline.
We believe these methods can potentially increase utility to the end-user by providing more information about model predictions retained in the model. Future work also includes improving the capability of the model to capture valence, which will likely improve the distribution performance as well. Additionally, we believe the method provides avenues into studying how emotion models can predict specific to groups of annotators or leverage the knowledge of annotators to improve zero-shot cross-corpus performance. 

\section{Acknowledgements}
This material is based in part upon work supported by the National Science Foundation (NSF IIS-RI 2230172 and IIS-RI 2230172) and National Institutes of Health (NIH R01MH130411).

\bibliographystyle{IEEEtran}
\bibliography{mybib}

\end{document}

%% file: tables/improv_results.tex
\begin{table}[t]
\centering
\setlength{\tabcolsep}{2pt}
\caption{MSP-Improv probability distribution results (*=statistical significant improvement compared to baseline, $\dagger$=statistical significant decline). $\uparrow$ indicates higher is better, $\downarrow$ indicates lower is better. Each metric's best result is bolded.}
\label{tab:improv-results}
\begin{tabular}{@{}lllll@{}}
\toprule
\textbf{Model} & \textbf{\begin{tabular}[c]{@{}l@{}}TVD$\downarrow$\end{tabular}} & \textbf{\begin{tabular}[c]{@{}l@{}}JSD$\downarrow$\end{tabular}} & \textbf{\begin{tabular}[c]{@{}l@{}}Activation\\CCC$\uparrow$\end{tabular}} & \textbf{\begin{tabular}[c]{@{}l@{}}Valence\\CCC$\uparrow$\end{tabular}} \\ \midrule
Baseline& .515±.004& .213±.003& .673±.008& .573±.020\\
MT& \textbf{.503±.001*}& \textbf{.211±.001}& \textbf{.741±.005*}& .571±.005\\
One-hot& .518±.006& .228±.004$\dagger$& .689±.014*& \textbf{.607±.017*}\\
\bottomrule
\end{tabular}
\vspace{-12pt}
\end{table}

%% file: tables/cc_results.tex
\begin{table}[]
\centering
\setlength{\tabcolsep}{2pt}
\caption{Cross-corpus zero-shot  \textbf{Act}ivation, \textbf{Val}ence results, *,$\dagger$,$\uparrow$,$\downarrow$ as in Table~\ref{tab:improv-results}. P: MSP-Podcast, I: IEMOCAP, M: MuSE}
\label{tab:cc-results}
\begin{tabular}{@{}cc|lll@{}}
\toprule
 & \multicolumn{1}{r}{\textbf{Dataset}} & \textbf{Baseline} & \textbf{MT-1} & \textbf{MT-12} \\ \midrule
\multirow{3}{*}{\textbf{TVD$\downarrow$}} & \textbf{P} & 0.601±0.003 & \textbf{0.507±0.002*} & 0.518±0.005* \\
 & \textbf{I} & 0.633±0.002 & 0.614±0.002* & \textbf{0.613±0.002*} \\
 & \textbf{M} & 0.530±0.004 & 0.484±0.007* & \textbf{0.470±0.002*} \\\midrule
\multirow{3}{*}{\textbf{JSD$\downarrow$}} & \textbf{P} & 0.274±0.002 & \textbf{0.213±0.001*} & 0.220±0.003* \\
 & \textbf{I} & 0.310±0.002 & \textbf{0.302±0.002*} & \textbf{0.302±0.002*} \\
 & \textbf{M} & 0.218±0.003 & 0.192±0.005* & \textbf{0.182±0.002*} \\\midrule
\multirow{3}{*}{\textbf{\begin{tabular}[c]{@{}c@{}}Act.\\ CCC$\uparrow$\end{tabular}}} & \textbf{P} & \textbf{0.261±0.014} & \textbf{0.261±0.008} & 0.235±0.012$\dagger$ \\
 & \textbf{I} & 0.374±0.010 & \textbf{0.429±0.010*} & 0.381±0.015 \\
 & \textbf{M} & 0.173±0.022 & 0.202±0.014 & \textbf{0.209±0.012*} \\\midrule
\multirow{3}{*}{\textbf{\begin{tabular}[c]{@{}c@{}}Val.\\ CCC$\uparrow$\end{tabular}}} & \textbf{P} & \textbf{0.368±0.003} & 0.332±0.009$\dagger$ & 0.302±0.011$\dagger$ \\
 & \textbf{I} & \textbf{0.321±0.011} & 0.255±0.007$\dagger$ & 0.219±0.011$\dagger$ \\
 & \textbf{M} & 0.198±0.017 & \textbf{0.202±0.013} & 0.162±0.007$\dagger$ \\ \bottomrule
\end{tabular}
\vspace{-18pt}
\end{table}